\documentclass[showkeys,eqsecnum,twocolumn,showpacs,preprintnumbers,amssymb,aps]{revtex4}
\usepackage{graphicx}
\usepackage{dcolumn}
\usepackage{amsthm,amsmath}

\begin{document}
\preprint{\today}

\title{Transition Properties of Low Lying States in Atomic Indium}
\vspace{0.5cm}

\author{B. K. Sahoo \footnote{Email: bijaya@prl.res.in}}
\affiliation{Theoretical Physics Division, Physical Research Laboratory, Ahmedabad-380009, India}
\author{B. P. Das}
\affiliation{Theoretical Astrophysics Group, Indian Institute of Astrophysics, Bangalore-560034, India}

\date{Received date; Accepted date}

\vskip1.0cm

\begin{abstract}
\noindent
We present here the results of our relativistic many-body calculations of 
various properties of the first six low-lying excited states of indium.
The calculations were performed using the relativistic coupled-cluster
method in the framework of the singles, doubles and partial triples 
approximation. We obtain a large lifetime $\sim 10s$ for the 
$[4p^6]5s^25p_{3/2}$ state, which had not been known earlier. Our 
results could be used to shed light on the reliability of the
lifetime measurements of the excited states of atomic indium that we have 
considered in the present work.
\end{abstract} 
\maketitle

\section{Introduction}
Indium (In) has been laser cooled and trapped a few years ago  \cite{laser}. 
Following this experiment, a proposal has 
been made to search for the permanent electric dipole moment (EDM) in this atom
\cite{pandey}. It would ineed be desirable to carry out high precision 
measurements and many-body calculations of other properties of this atom.  
A few measurements of the magnetic dipole hyperfine structure constants
of the first three low-lying states of In are already
available \cite{eck,george}. However, the reported theoretical results 
obtained using different variants of the relativistic coupled-cluster (RCC) 
method at the singles, 
doubles and important triples excitations level (CCSD(T) method) are not able
to reproduce them to within one percent accuracy \cite{pandey,safronova}. This
suggests that the role of correlation effects for this property is of crucial
importance. In addition, it would also be worthwhile to calculate different
transition amplitues in In for a number of reasons. First of all, 
the behavior of the correlation effects in these properties could be
quite different than in the hyperfine structure constants. Furthermore,
these amplitudes in conjunction with the hyperfine constants can be employed 
to verify the accuracy of the wave functions for the proposed EDM calculations 
\cite{pandey} or perhaps for the parity nonconservation in this system if at 
all an experiment to observe this effect can be carried out in this atom and 
also to determine the polarizabilities, lifetimes, oscillator strengths, 
branching ratios etc. for various states.

\begin{center}
\begin{figure}[h]
\includegraphics[width=12.0cm,clip=true]{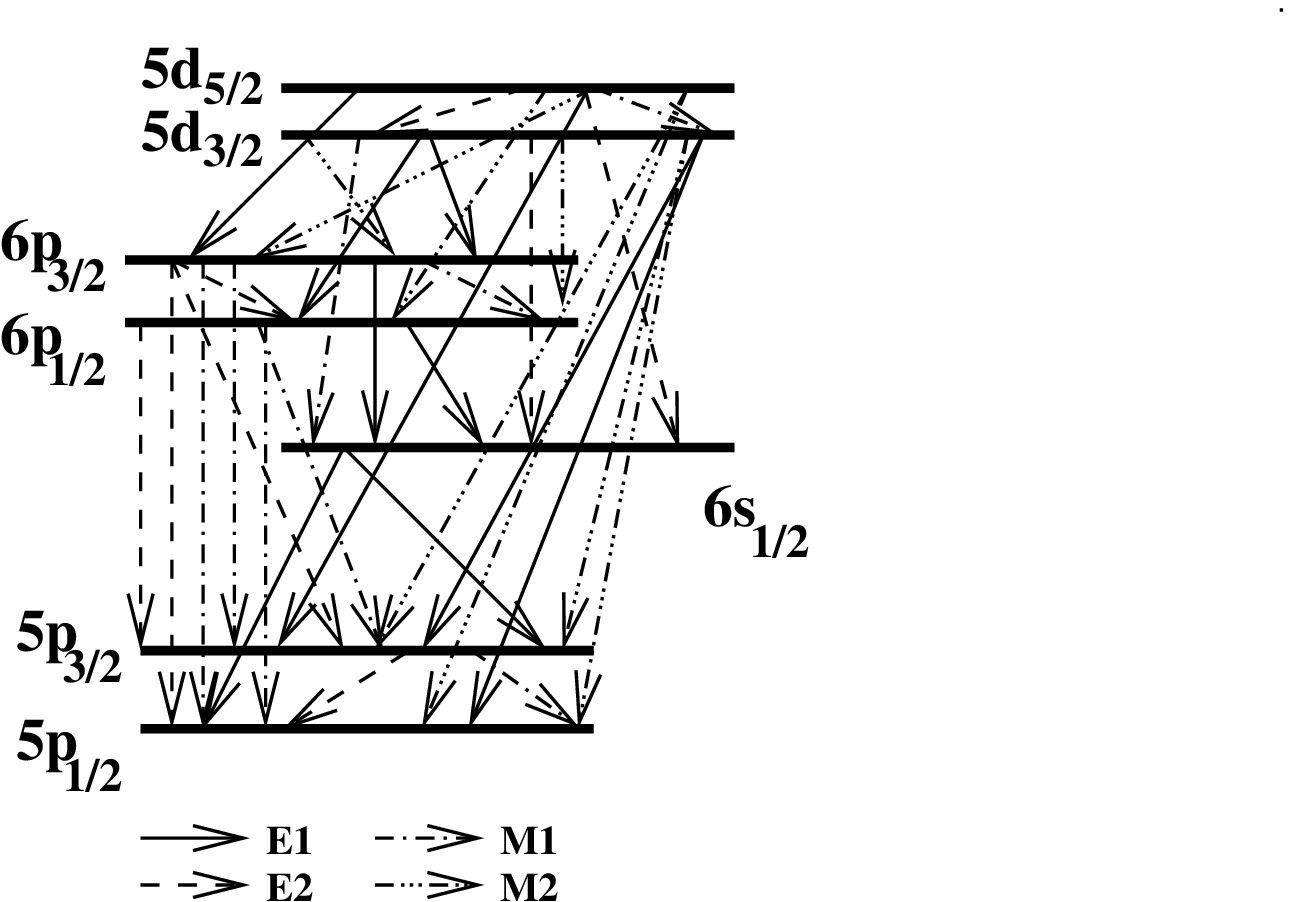}
\caption{Schematic low-lying energy level diagrams and decay channels of the low-lying states in In.}
\label{fig1}
\end{figure}
\end{center}
 In this work, we calculate the excitation energies (EEs) and different 
transition 
amplitudes due to allowed and forbidden electromagnetic transitions among the 
first six low-lying states, giving a total of 34 possible 
transitions (see Fig. \ref{fig1}), using the relativistic CCSD(T) method. 
These results are further used to determine transition rates, branching
ratios and lifetimes of the above states. These properties are also important
from an astrophysical point of view \cite{vitas,jaschek}. Safronova {\it et al.}
have reported EEs and electric dipole (E1) transition amplitudes for a number
of states and compared with previous calculations and measurements
\cite{safronova}. However, they have only considered these transition
amplitudes to estimate the lifetimes of various states, but contributions from 
the forbidden transitions are not included. In our calculations, we have taken
into account the forbidden transition amplitudues in the 
evaluation of the lifetimes of different states.

\section{Theory and Method of calculations}
The transition rates (in $s^{-1}$) due to various transitions are 
given by \cite{shore}
\begin{eqnarray}
A^{\text{E1}}_{f \rightarrow i} &=& \frac {2.02613 \times 10^{18} }{\lambda^3 (2J_f+1)} S_{f \rightarrow i}^{\text{E1}},
\label{eqn1} \\
A^{\text{M1}}_{f \rightarrow i} &=& \frac {2.69735 \times 10^{13}}{\lambda^3 (2J_f+1)} S_{f \rightarrow i}^{\text{M1}},
\label{eqn2} \\
A^{\text{E2}}_{f \rightarrow i} &=& \frac {1.11995 \times 10^{18}}{\lambda^5 (2J_f+1)} S_{f \rightarrow i}^{\text{E2}},
\label{eqn3} \\
\text{and} \ \ \ \ \ \ \ \ \ \ &&  \nonumber \\
A^{\text{M2}}_{f \rightarrow i} &=& \frac {1.491 \times 10^{13}}{\lambda^5 (2J_f+1)} S_{f \rightarrow i}^{\text{M2}},
\label{eqn4} 
\end{eqnarray}
where $\lambda$ (in $\AA$) and $S_{f \rightarrow i}^{\text{O}} (= |\langle f || \text{O} || i \rangle|^2)$ (in atomic unit (au)) are the wavelengths and line strengths due to the 
corresponding transition operator $\text{O}$, respectively.

The lifetime ($\tau_f$) of a given state $f$ is just the reciprocal of the 
total transition
rate of that state  due to all possible transition channels; i.e.
\begin{eqnarray}
\tau_f &=& \frac{1} {\sum_{\text{O},i} A^{\text{O}}_{f \rightarrow i}},
\label{eqn5} 
\end{eqnarray}
where $A^{\text{O}}_{f \rightarrow i}$ is the transition rate due to
operator $\text{O}$ and sum over $i$ and $\text{O}$ represents the total transition
rate from state $f$ to all possible states $i$ and due to all possible
operators.

The branching ratios due to an operator $\text{O}$ from a state $f$ due to the lower 
states are given by
\begin{eqnarray}
\Gamma^{\text{O}}_{f \rightarrow i} &=& \frac {A^{\text{O}}_{f \rightarrow i}} {\sum_{\text{O},i} A^{\text{O}}_{f \rightarrow i}} \nonumber \\ 
 &=& \tau_f A^{\text{O}}_{f \rightarrow i}.
\label{eqn6} 
\end{eqnarray}

To evaluate the line strengths, we use the following reduced matrix 
elements at the single particle orbitals level for the E1, M1, E2 and M2 
operators \cite{johnson}
\begin{widetext}
\begin{eqnarray}
\langle \kappa_f\, ||\, e1\, ||\, \kappa_i \rangle
&=&\langle \kappa_f \,||\, C^{(1)} \,||\, \kappa_i \rangle 
\int_0^{\infty} dr r \{ (P_f P_i + Q_f Q_i) - \frac{\omega  r}{5 \alpha} [ \frac{\kappa_f - \kappa_i}{2} (P_f Q_i + Q_f P_i) + (P_f Q_i - Q_f P_i) ] \},
\label{eqn7}\\
\langle \kappa_f\, ||\,m1\,||\, \kappa_i \rangle &=&  \langle - \kappa_f\, ||\,C^{(1)}\,||\,\kappa_i \rangle \int_0^{\infty} dr r \frac {(\kappa_f+\kappa_i)}{\alpha} (P_fQ_i+Q_fP_i),
\label{eqn8}\\
\langle \kappa_f\, ||\,e2\,||\,\kappa_i \rangle &=& \langle \kappa_f\, ||\,C^{(2)}\,||\, \kappa_i \rangle \int_0^{\infty} dr r^2 \{ (P_fP_i+Q_fQ_i) 
- \frac{ \omega r}{7 \alpha} [ \frac {\kappa_f-\kappa_i} {3} (P_fQ_i+Q_fP_i)
+ (P_fQ_i-Q_fP_i)] \},
\label{eqn9}\\
\text{and} \ \ \ \ \ \ \ \ \ \ \ \ \ \ && \nonumber \\
\langle \kappa_f\, ||\,m2\,||\,\kappa_i \rangle &=& \langle - \kappa_f\, ||\,C^{(2)}\,||\, \kappa_i \rangle \int_0^{\infty} dr r^2 \frac {(\kappa_f+\kappa_i)}{3 \alpha} (P_fQ_i+Q_fP_i),
\end{eqnarray}
\label{eqn10}
\end{widetext}
where, $j's$  and  $\kappa's$ are the orbital and relativistic angular momentum
quantum numbers, respectively, $P$ and $Q$ represent the radial parts of 
large and small components of single particle Dirac orbitals, respectively,
$\omega= \epsilon_f - \epsilon_i$ for the orbital energies $\epsilon$s, 
$\alpha$ is the fine structure constant and the reduced Racah coefficients are given by
\begin{eqnarray}
\langle \kappa_f\, ||\, C^{(k)}\,||\, \kappa_i \rangle &=& (-1)^{j_f+1/2} \sqrt{(2j_f+1)(2j_i+1)} \ \ \ \ \ \ \ \ \nonumber \\
                  &&        \left ( \begin{matrix} 
                              j_f & k & j_i \\
                              1/2 & 0 & -1/2 \\
                                       \end{matrix}
                            \right ) \pi(l_{\kappa_f},k,l_{\kappa_i}), \ \ \ \ \
\label{eqn11}
\end{eqnarray}
with
\begin{eqnarray}
  \pi(l,m,l') &=&
  \left\{\begin{array}{ll}
      \displaystyle
      1 & \mbox{for } l+m+l'= \mbox{even}
         \\ [2ex]
      \displaystyle
        0 & \mbox{otherwise.}
    \end{array}\right.
\label{eqn12}
\end{eqnarray}
In the above expressions and in the remaining part of the paper, we have 
used au unless they are mentioned explicitly.

In order to determine the above properties, we calculate the atomic wave function 
($|\Psi_v \rangle$) with a valence orbital $v$ by expressing it in the RCC framework as
\begin{eqnarray}
|\Psi_v \rangle &=& e^T \{1+S_v\} |\Phi_v \rangle , 
\label{eqn13}
\end{eqnarray}
where we define a reference state $|\Phi_v \rangle$ by appending the appropriate
valence orbital $v$ to the Dirac-Fock (DF) wave function ($|\Phi_0\rangle$) with the
configuration similar to cadmium; i.e. $[4p^6]4d^{10}5s^2$. Here $T$ and $S_v$
represent the excitation operators due to core-core and core-valence electron
correlations. In the CCSD(T) method, the $T$ and $S_v$ operators are defined as
\begin{eqnarray}
T = T_1 + T_2 
&\text{and}& 
S_v =  S_{1v} + S_{2v} , 
\label{eqn14}
\end{eqnarray}
\noindent
where $1$ and $2$ in the subscripts represent for single and double excitations, respectively. 

The equations determining the coupled-cluster amplitudes and energy can be 
expressed in compact forms as
\begin{eqnarray}
\langle \Phi_0^L |\{\widehat{He^T}\}|\Phi_0 \rangle &=& \delta_{0,L} \Delta E_{corr} 
\label{eqn15}
\end{eqnarray}
and
\begin{eqnarray}
\langle \Phi_v^L|\{\widehat{He^T}\} \{1+S_v\}|\Phi_v\rangle &=& \langle \Phi_v^L|1+S_v|\Phi_v\rangle \nonumber \\ && \langle \Phi_v|\{\widehat{He^T}\} \{1+S_v\} |\Phi_v\rangle \nonumber \\
 &=& \langle \Phi_v^L|\delta_{L,v}+S_v|\Phi_v\rangle \Delta E_v, \ \ \ \ \ \ \ \
\label{eqn16}
\end{eqnarray}
where the superscript $L (=1,2)$ represents for the excited hole-paerticle
states, $\widehat{He^T}$ denotes the connected terms of the  
Dirac-Coulomb (DC) Hamiltonian with the $T$ operators, $\Delta E_{corr}$ and
$\Delta E_v$ are the correlation energy and attachment energy (also equivalent
to negative of the ionization potential (IP)) of the electron of orbital $v$, respectively.
The reference states in Eq. (\ref{eqn15}) and 
Eq. (\ref{eqn16}) contain different number of particles, hence the 
Hamiltonian used in the respective equations describe different number
of particles in our Fock space representation. Contributions from the important valence triple excitations
are included perturbatively through the above equations.

The transition matrix element of a physical operator $\text{O}$ between 
$|\Psi_f \rangle$ and $|\Psi_i \rangle$ in our approach is given by
\begin{widetext}
\begin{eqnarray}
\frac{\langle \Psi_f | \text{O} | \Psi_i \rangle}{\sqrt{\langle \Psi_f|\Psi_f\rangle}\sqrt{\langle \Psi_i|\Psi_i\rangle}} &=& \frac{\langle \Phi_f | \{ 1+ S_f^{\dagger}\} \overline{ \text{O}} \{ 1+ S_i\} |\Phi_i\rangle}{\sqrt{\langle \Phi_f | \overline{N} + S_f^{\dagger} \overline{N} S_f |\Phi_f\rangle}\sqrt{\langle \Phi_i | \overline{N} + S_i^{\dagger} \overline{N} S_i |\Phi_i\rangle}},
\label{eqn17}
\end{eqnarray}
\end{widetext}
where $\overline{ \text{O}}=e^{T^{\dagger}} \text{O} e^T$ and $\overline{N}=e^{T^{\dagger}} e^T$ are two non-truncating series in the above expression. We evaluate them by considering terms whose leading contributions arise
in fourth order perturbation theory or lower. Contributions from to the normalization of
the wave functions (${\cal N}$) are determined explicitly as follows
\begin{eqnarray}
{\cal N} &=& \frac{\langle \Psi_f | \text{O} | \Psi_i \rangle}{\sqrt{\langle \Psi_f|\Psi_f\rangle}\sqrt{\langle \Psi_i|\Psi_i\rangle}} - \langle \Psi_f | \text{O} | \Psi_i \rangle \ \ \ \ \ \ \ \nonumber \\
 &=& \langle \Psi_f | \text{O} | \Psi_i \rangle \left [ \frac{1}{\sqrt{\langle \Psi_f|\Psi_f\rangle}\sqrt{\langle \Psi_i|\Psi_i\rangle}} - 1 \right ]. \ \ \ \ \ \ 
\label{eqn18}
\end{eqnarray}

We have used Gaussian type orbitals (GTOs) to construct the single particle
orbitals for the Dirac-Fock ($|\Phi_0 \rangle$) wave function. 
The large and small components of the Dirac orbitals in this case are
expressed as
\begin{eqnarray}
P_{\kappa}(r) &=& \sum_k c_k^P r^{l_{\kappa}} e^{-\alpha_k r^2}
\label{eqn19}
\end{eqnarray}
and
\begin{eqnarray}
Q_{\kappa}(r) &=& \sum_k c_k^Q r^{l_{\kappa}} \left ( \frac{d}{dr} + \frac{\kappa}{r} \right ) e^{-\alpha_k r^2}, 
\label{eqn20}
\end{eqnarray}
where the summation over $k$ is for total number of GTOs used in each symmetry, 
$c_k^P$ and $c_k^Q$ are the normalization constants for the large and
small components, respectively, and we use $(\frac{d}{dr}+\frac{\kappa}{r})$ 
operator to expand the small component Dirac orbitals to maintain the kinetic
balance condition with its large component. In the present calculations, we
have considered 9 relativistic symmetries (up to $g$ symmetry) and 28 GTOs for each symmetry to generate the orbitals. In order to optimise
the exponents to describe orbitals from various symmetries in a smooth manner,
we use the even tempering condition
\begin{equation}
\alpha_k = \alpha_0 \beta^{k-1}, 
\label{eqn21}
\end{equation}
where $\alpha_0$ and $\beta$ are two arbitary parameters that can be chosen 
suitably for different symmetries. We have considered $\alpha_0=7.5\times10^{-4}$ for all the symmetries and $\beta$ are taken as $2.53$, $2.45$, $2.58$, 
$2.75$ and $2.83$ for $s$, $p$, $d$, $f$ and $g$ orbitals, respectively. For
the RCC calculations, we have considered excitations up to first $16s$, $16p$, 
$16d$, $14f$ and $13g$ orbitals as the remaining orbitals have large continuum
energies.

\section{Results and Discussion}
In accordance with Koopman's theorem, the energies of the virtual orbitals 
obtained in our calculations are the IPs at the DF level, since our DF wave
function is computed using the closed-shell configuration $[4p^6]4d^{10}5s^2$.
We present the IPs from NIST \cite{nist}, from other calculation as well as our
calculations in Table \ref{tab1}.
\begin{table}[h]
\caption{IPs ($\Delta E_v$s) of different states of In in cm$^{-1}$. Absolute error of our CCSD(T) results compared to the quoted results in \cite{nist} are given as $\Delta$.}
\begin{ruledtabular}
\begin{center}
\begin{tabular}{lccccc}
State  &  $^a$NIST & $^b$Others & $^c$Koopman & $^c$CCSD(T) & $\Delta$ \\
  &  in $cm^{-1}$ & in $cm^{-1}$ & in $cm^{-1}$ & in $cm^{-1}$ & in \% \\ 
\hline\\
$5p_{1/2}$ & 46670.11 & 46189 & 41521.74 & 46581.47 & $0.19$ \\   
$5p_{3/2}$ & 44457.51 & 44031 & 39522.20 & 44361.04 & $0.22$ \\
$6s_{1/2}$ & 22297.15 & 22442 & 20567.70 & 22291.74 & $0.02$ \\   
$6p_{1/2}$ & 14853.21 & 14833 & 13977.92 & 14819.07 & $0.23$ \\   
$6p_{3/2}$ & 14554.89 & 14532 & 13718.04 & 14519.46 & $0.24$ \\
$5d_{3/2}$ & 13777.90 & 13581 & 12389.68 & 13633.48 & $1.05$ \\
$5d_{5/2}$ & 13754.57 & 13554 & 12373.64 & 13603.88 & $1.10$ \\
\end{tabular}
\end{center}
\end{ruledtabular}
\label{tab1}
$^a$ Reference \cite{nist}.\\
$^b$ Reference \cite{safronova}.\\
$^c$ This work.
\end{table}

As observed from Table \ref{tab1}, our IP results are within $0.5$\% 
except for the $5d$ states (which are around 1\% accurate) compared with the
results given in \cite{nist}. In an earlier work, Safronova {\it et al.}
have reported the results for these quantities based on the linearlized version of
the relativistic CCSD (T) method using a B-spline basis \cite{safronova}. 
The major differences between this and our work are the different basis sets
used in the two calculations and the additional non linear clusters in our
calculation. Our results are in better agreement with the high precision NIST 
results than those of \cite {safronova} for all the states that we have
considered. Given the high accuracy of our IPs and therefore the excitation energies, we can 
accurately determine the wavelengths
for various transitions in order to determine the
transition rates and the lifetimes of different excited states. We can also use the 
wavelengths from NIST data to obtain the lifetimes and compare them with the
results from the relativistic CCSD(T) method.
\begin{table}[h]
\caption{Line strengths (in au) due to allowed and forbidden transitions 
between different states in In. Numbers given in the parentheses and square 
brackets represent estimated errors and powers in 10, respectively.}
\begin{ruledtabular}
\begin{center}
\begin{tabular}{lccc}
Transition  &  DF & CCSD(T) & Others \cite{safronova} \\
\hline\\
$5d_{5/2} \xrightarrow{E1} 6p_{3/2}$ & 251.95 & 188(2) & 186 \\
$5d_{5/2} \xrightarrow{M2} 6p_{3/2}$ & $4.9[3]$ & $3.8(1)[3]$ & \\
$5d_{5/2} \xrightarrow{M2} 6p_{1/2}$ & $893.6$ & $685(5)$ \\
$5d_{5/2} \xrightarrow{E1} 5p_{3/2}$ & $20.55$ & $16.3(5)$  & 15.2 \\
$5d_{5/2} \xrightarrow{M2} 5p_{3/2}$ & $399.59$ & $474(10)$ &  \\
$5d_{5/2} \xrightarrow{M2} 5p_{1/2}$ & $62.44$ & $92(3)$ &  \\
$5d_{5/2} \xrightarrow{M1} 5d_{3/2}$ & 2.40 & 2.41(1) &  \\
$5d_{5/2} \xrightarrow{E2} 5d_{3/2}$ & $5.0[3]$ & $2.9(2)[3]$ &  \\
$5d_{5/2} \xrightarrow{E2} 6s_{1/2}$ & $9.4[3]$ & $6.6(1)[3]$ & \\
$5d_{3/2} \xrightarrow{E1} 6p_{3/2}$ & 27.88 & 20(1) & 20.5 \\
$5d_{3/2} \xrightarrow{M2} 6p_{3/2}$ & 0.0 & $3.5(1)[-3]$  & \\
$5d_{3/2} \xrightarrow{E1} 6p_{1/2}$ & $139.07$ & $104(5)$ & 103 \\
$5d_{3/2} \xrightarrow{M2} 6p_{1/2}$ & $83.44$ & $65(3)$ & \\
$5d_{3/2} \xrightarrow{E1} 5p_{3/2}$ & $2.30$ & $1.84(2)$ & 1.71 \\
$5d_{3/2} \xrightarrow{M2} 5p_{3/2}$ & 0.0 & 0.27(1) & \\
$5d_{3/2} \xrightarrow{E1} 5p_{1/2}$ & $9.84$ & $7.7(4)$ & 7.24 \\
$5d_{3/2} \xrightarrow{M2} 5p_{1/2}$ & $5.90$ & $4.9(3)$ &  \\
$5d_{3/2} \xrightarrow{M1} 6s_{1/2}$ & $4.8[-12]$ & $2(1)[-9]$ & \\
$5d_{3/2} \xrightarrow{E2} 6s_{1/2}$ & $6.3[3]$ & $4.4(1)[3]$ &  \\
$6p_{3/2} \xrightarrow{M1} 6p_{1/2}$ & $1.33$ & $1.33(1)$ &  \\
$6p_{3/2} \xrightarrow{E2} 6p_{1/2}$ & $1.64[4]$ & $1.32(1)[4]$ &  \\
$6p_{3/2} \xrightarrow{E1} 6s_{1/2}$ & 88.96 & 72.9(1) & 70.3 \\
$6p_{3/2} \xrightarrow{M1} 5p_{3/2}$ & $4.9[-9]$ & $2.1(5)[-4]$& \\
$6p_{3/2} \xrightarrow{E2} 5p_{3/2}$ & 131.93 & 106(8) & \\
$6p_{3/2} \xrightarrow{M1} 5p_{1/2}$ & $9.2[-4]$ & $6(1)[-4]$ &  \\
$6p_{3/2} \xrightarrow{E2} 5p_{1/2}$ & $96.90$ & $77.3(8)$ &  \\
$6p_{1/2} \xrightarrow{E1} 6s_{1/2}$ & $45.81$ & $37.5(1)$ & 36.1 \\
$6p_{1/2} \xrightarrow{M1} 5p_{3/2}$ & $1.0[-3]$ & $1.4(3)[-3]$ &  \\
$6p_{1/2} \xrightarrow{E2} 5p_{3/2}$ & $149.33$ & $120(7)$ &  \\
$6p_{1/2} \xrightarrow{M1} 5p_{1/2}$ & $3.9[-10]$ & $1.2(8)[-5]$ &  \\
$6s_{1/2} \xrightarrow{E1} 5p_{3/2}$ & $11.26$ & $8.8(2)$ & 8.56 \\
$6s_{1/2} \xrightarrow{E1} 5p_{1/2}$ & $4.68$ & $3.67(2)$ & 3.64 \\
$5p_{3/2} \xrightarrow{M1} 5p_{1/2}$ & $1.33$ & $1.31(1)$ & \\
$5p_{3/2} \xrightarrow{E2} 5p_{1/2}$ & $236.42$ & $181(1)$ & \\
\end{tabular}
\end{center}
\end{ruledtabular}
\label{tab2}
\end{table}

In Table \ref{tab2}, we present the line strengths obtained using the DF and relativistic
CCSD(T) methods for both the allowed and fobidden transitions. Safronova 
{\it et al.} have given the results only for the allowed transitions
\cite{safronova} and they have not verified explicitly the contributions from the
forbidden transitions. In fact, our transition strengths for the allowed
transitions differ slightly from theirs and the cause of the
differences between the approximations employed in the two cases
have already been discussed earlier. The line strengths due to the forbidden 
transitions are not small in many of the cases. The influence of
the forbidden transitions should be verified in the determination of transition rates, BRs and lifetime estimations as they may be important in some cases.
\begin{table*}[t]
\caption{Wavelengths ($\lambda$) in $\AA$, transition rates ($A$) in $s^{-1}$,
branching ratios ($\Gamma$) and lifetimes ($\tau$) in $ns$ for the considered
excited states in In. We consider the calculated and experimental values of 
$\lambda$s to determine the above quantities which are given as I and II,
respectively. We present the recommended (Reco) values for $\tau$s after
accounting possible errors and compared them with their experimental (Expt)
results. Numbers given in the parentheses and square brackets represent 
estimated errors and powers in 10, respectively.}
\begin{ruledtabular}
\begin{center}
\begin{tabular}{lccccccccccc}
Upper & Lower & Channel & \multicolumn{2}{c}{$\lambda_{f \rightarrow i}$} & \multicolumn{2}{c}{$A^{\text{O}}_{f \rightarrow i}$} & $\Gamma^{\text{O}}_{f \rightarrow i}$ & \multicolumn{3}{c}{$\tau_f$} & $\tau_f$\\
 state ($f$) &  state ($i$) & O & I & II & I & II & & I & II & Reco & Expt \\ 
\hline\\
$5d_{5/2}$ & $6p_{3/2}$ & E1 & $1.09[5]$ & $1.25[5]$ & 48781.17 & 32544.0 & 0.0002 & 6.22 & 6.27 & 6.3(2) & $^a$7.6(5) \\
           & $6p_{3/2}$ & M2 &   & & $6.1[-10]$ & $3.1[-10]$ & $\sim0.0$ &  & & & $^b$7.1(6)\\
           & $6p_{1/2}$ & M2 & $8.23[4]$ & $9.10[4]$ & $4.5[-5]$ & $2.7[-5]$ & $\sim0.0$ & \\
           & $5p_{3/2}$ & E1 & 3251 & 3257 & $1.61[8]$ & $1.60[8]$ & 0.9998 &  & \\
           & $5p_{3/2}$ & M2 &  & & $3.2[-3]$ & $3.2[-3]$ & $\sim0.0$ & & \\
           & $5p_{1/2}$ & M2 & 3032 & 3038 &  $8.9[-4]$ & $8.8[-4]$ & $\sim0.0$ & \\
           & $5d_{3/2}$ & M1 & $3.38[6]$ & $4.29[6]$ & $2.8[-7]$ & $1.4[-7]$ & $\sim0.0$ & \\
           & $5d_{3/2}$ & E2 & & & $1.2[-12]$ & $3.7[-12]$ & $\sim0.0$ & & \\
           & $6s_{1/2}$ & E2 & $1.15[4]$ & $1.14[4]$ & 6.10 & 5.61 & $\sim0.0$ & \\
$5d_{3/2}$ & $6p_{3/2}$ & E1 & $1.13[5]$ & $1.29[5]$ & 7316.78 & 4935.01 & $3.0[-5]$ & 5.99 & 6.03 & 6.0(3) & $^a$6.3(5) \\
           & $6p_{3/2}$ & M2 &  & & $7.1[-16]$ & $3.7[-16]$ & $\sim0.0$ & & & & $^c$7.0(4) \\
           & $6p_{1/2}$ & E1 & $8.43[4]$ & $9.30[4]$ & $8.78[4]$ & $6.55[4]$ & $4.0[-4]$ \\
           & $6p_{1/2}$ & M2 & & $5.7[-11]$ & $3.5[-11]$ &  $\sim0.0$\\
           & $5p_{3/2}$ & E1 & 3254 & 3259 & $2.71[7]$ & $2.69[7]$ & 0.16 \\
           & $5p_{3/2}$ & M2 &  &  & $2.8[-6]$ & $2.7[-6]$ &  $\sim0.0$\\
           & $5p_{1/2}$ & E1 & 3035 & 3040 & $1.40[8]$ & $1.39[8]$ & 0.84 \\
           & $5p_{1/2}$ & M2 &  & $7.1[-5]$ & $7.1[-5]$ &  $\sim0.0$ \\
           & $6s_{1/2}$ & M1 & $1.16[4]$ & $1.17[4]$ & $1.1[-8]$ & $1.0[-8]$ &  $\sim0.0$\\
           & $6s_{1/2}$ & E2 &  &  & 5.99 & 5.53 & $\sim0.0$\\
$6p_{3/2}$ & $6p_{1/2}$ & M1 & $3.34[5]$ & $3.35[5]$ & $2.4[-4]$ & $2.4[-4]$ & $\sim0.0$ & 57.67 & 58.34 & 58(1) & $^d$55(4)\\
           & $6p_{1/2}$ & E2 &        &        &  $8.9[-7]$ & $8.7[-7]$ & $\sim0.0$ & \\
           & $6s_{1/2}$ & E1 & $1.29[4]$ & $1.29[4]$ & $1.73[7]$ & $1.71[7]$ & 0.9999 & \\
           & $5p_{3/2}$ & M1 & 3351 & 3344 & 0.04 & 0.04 & $\sim0.0$ & \\
           & $5p_{3/2}$ & E2 &  &  & 69.99 & 70.73 & $4.1[-6]$ & \\
           & $5p_{1/2}$ & M1 & 3119 & 3114 & 0.13 & 0.13 & $\sim0.0$ & \\
           & $5p_{1/2}$ & E2 &  &  & 73.34 & 73.93 & $4.3[-6]$ &  & \\
$6p_{1/2}$ & $6s_{1/2}$ & E1 & $1.34[4]$ & $1.34[4]$ & $1.58[7]$ & $1.56[7]$ & 0.9999 & 63.16 & 63.81 & 63.8(8) & $^d$55(4) \\
           & $5p_{3/2}$ & M1 & 3385 & 3378 & 0.49 & 0.49 & $\sim0.0$ & \\
           & $5p_{3/2}$ & E2 &  &  & 150.79 & 152.35 & $9.7[-6]$ \\
           & $5p_{1/2}$ & M1 & 3148 & 3143 & 0.005 & 0.005 & $\sim0.0$ & \\
$6s_{1/2}$ & $5p_{3/2}$ & E1 & 4531 & 4513 & $9.58[7]$ & $9.70[7]$ & 0.6431 & 6.71 & 6.63 & 6.6(2) & $^a$7.5(7) \\
           & $5p_{1/2}$ & E1 & 4117 & 4103 & $5.33[7]$ & $5.38[7]$ & 0.3569 & & & & $^e$7.0(3) \\
           & & & & &  &  &  & & & & $^f$7.4(3) \\
$5p_{3/2}$ & $5p_{1/2}$ & M1 & 4.50[4] & 4.52[4] & 0.0967 & 0.0957 & 0.9969 & $10.31[9]$ & $10.42[9]$ & $10.4(2)[9]$ & \\
           & $5p_{1/2}$ & E2 &  &  & 0.0003 & 0.0003 & 0.0031 & \\
\end{tabular}
\end{center}
\end{ruledtabular}
\label{tab3}
$^a$Reference \cite{andersen}. \\
$^b$Reference \cite{zimmermann}. \\
$^c$Reference \cite{brieger}. \\
$^d$Reference \cite{ewiss}. \\
$^e$Reference \cite{norton}. \\
$^f$Reference \cite{havey}.
\end{table*}

We present the wavelengths, transition rates, branching ratios and lifetimes 
of different states of In in Table \ref{tab3}. These quantities are
determined using both the calculated wavelengths that are estimated from the
excitation energies obtained in this work and the experimental wavelengths from 
NIST data \cite{nist}. The {\it ab initio} results are given as I and wherever
the experimental wavelengths used are given as II. We also give measured
lifetimes results based on different experimental techniques \cite{andersen,
zimmermann,brieger,ewiss,norton,havey} in the same table.

The difference between the experimental results and obtained calculations
for EEs are treated as possible uncertainties associated with them, which
are given in percentage in Table \ref{tab1}. Uncertainties in the calculated transition
matrix elements are obtained by finding out the contributions from higher
angular momentum orbitals using the dominant many-body perturbation
diagrams and are mentioned in the parenthesis of the results presented in
Table \ref{tab2}. In the final lifetime estimation of various states, we 
consider central values given as II in Table \ref{tab3} and uncertainties
are determined from the above error bars. These results are reported as
recommended values (Reco) in Table \ref{tab3}.

As we find from the above table, the branching ratios due to the allowed
transition channels completely dominate over the forbidden transition 
channels. Therefore, the lifetimes of the excited states except for the 
$5p_{3/2}$ state are almost entirely determined by the allowed transitions. The
$5p_{3/2}$ state is the fine structure partner of the the ground state, its
lifetime is determined from the forbidden transitions. For this case, the M1 
transition
clearly dominates over the E2 transition and that is evident from their 
branching ratios. We obtain a large lifetime, $\sim10s$, for this state. The 
lifetime of the $5d_{5/2}$ state obtained from our calculation is in reasonable
agreement with the available experimental data. We find that the lifetime of 
this state is almost entirely due to the E1 decay channel to $5p_{3/2}$ state. 
As the wavelength of the transition $5d_{5/2} \rightarrow 6p_{3/2}$ is very 
large, the branching ratio of the $5d_{5/2}$ state is small. Our calculated
lifetime for the $5d_{3/2}$ state agrees well with the experimental results. We
find 84\% and 16\% branching ratios from this state to the ground and $5p_{3/2}$
states, respectively through the E1 channel. Contributions from the forbidden
transitions are also negligible in this case. Similar agreement between
our calculated and experimental results for the lifetimes of the $6p$ states
are found, but the experimental results have large error bars compared to our
calculations. There is also a marginal difference between the measured lifetimes
and the calculated lifetimes of the $6s$ state, although they are within the
common error bar. The branching ratios from this state to the ground and 
$5p_{3/2}$ states are of the order of 35\% and 64\%, respectively. This trend is different for the $5d_{3/2}$ state as discussed
above. 

\section{Conclusion}
We have estimated the branching ratios and lifetimes of certain low-lying excited 
states of indium. We have carried out calculations of the excitation energies
and line strengths using the relativistic coupled cluster method. We have
also compared our {\it ab initio} results with the results obtained using the
experimental wavelengths and measured lifetimes from different experimental 
techniques. We find that the forbidden transitions do not
contribute significantly to the lifetimes of most of the states that we have
considered. A large lifetime for the $5p_{3/2}$ state
($\sim10s$) has been found from this work, which is completely due to the
forbidden transitions to its fine structure partner; the ground state.

\section*{Acknowledgment}
These calculations were carried out using the PRL HPC 3TFLOP cluster at PRL 
and the CDAC Param Padma TeraFlop supercomputer.

\end{document}